\documentclass[12pt]{article}
\begin{document}
\begin{center}
{\large\bf Cosmic Microwave Background, Accelerating Universe and
Inhomogeneous Cosmology} \vskip 0.3 true in {\large J. W. Moffat}
\vskip 0.3 true in {\it The Perimeter Institute for Theoretical
Physics, Waterloo, Ontario, N2J 2W9, Canada} \vskip 0.3 true in
and \vskip 0.3 true in {\it Department of Physics, University of
Waterloo, Waterloo, Ontario N2Y 2L5, Canada}
\end{center}
%\date{\today}
\begin{abstract}%
We consider a cosmology in which a spherically symmetric large
scale inhomogeneous enhancement or a void are described by an
inhomogeneous metric and Einstein's gravitational equations. For a
flat matter dominated universe the inhomogeneous equations lead to
luminosity distance and Hubble constant formulas that depend on
the location of the observer. For a general inhomogeneous
solution, it is possible for the deceleration parameter to differ
significantly from the FLRW result. The deceleration parameter
$q_0$ can be interpreted as $q_0 > 0$ ($q_0=1/2$ for a flat matter
dominated universe) in a FLRW universe and be $q_0 < 0$ as
inferred from the inhomogeneous enhancement that is embedded in a
FLRW universe. A spatial volume averaging of local regions in the
backward light cone has to be performed for the inhomogeneous
solution at late times to decide whether the decelerating
parameter $q$ can be negative for a positive energy condition. The
CMB temperature fluctuations across the sky can be unevenly
distributed in the northern and southern hemispheres in the
inhomogeneous matter dominated solution, in agreement with the
analysis of the WMAP power spectrum data by several authors. The
model can possibly explain the anomalous alignment of the
quadrupole and octopole moments observed in the WMAP data.
\end{abstract}
\vskip 0.2 true in e-mail: john.moffat@utoronto.ca
%\pacs{ }

\section{Introduction}

The problem of how to explain the accelerating expanding universe
given the SNe Ia supernova data~\cite{Perlmutter,Riess} in
combination with the WMAP data has led to a host of solutions,
ranging from modified gravity theories to a quintessence field and
a cosmological constant~\cite{Peebles,Straumann}. In the following
we shall take the position that the homogeneous and isotropic
Friedmann-Lema\^{i}tre-Robertson-Walker (FLRW) cosmology is the
most symmetric model which describes many of the basic features of
the data. However, important physical features may not be
explained by this first approximation and a more complete
description is given by an inhomogeneous
cosmology~\cite{Krasinski}. We shall investigate a universe in
which a spherically symmetric inhomogeneous enhancement or void is
embedded in an asymptotic FLRW universe. We find that the
inhomogeneities permitted by observation can lead to a
reinterpretation of the luminosity distance $d_L$ of a
cosmological source in terms of its red shift $z$. The time
evolution and the expansion rate of the inhomogeneous universe can
lead to intrinsic effects such as cosmic variance at large angles,
long-wavelength perturbations not described by a FLRW homogeneous
and isotropic universe and late-time nonlinear effects. It is
possible that the observed acceleration of the universe is caused
by late-time inhomogeneity associated with large scale
structure~\cite{Moffat}. This is important, for it is difficult to
explain theoretically the postulated dark energy that causes the
acceleration of the universe. Tomita has investigated the
interpretation of the luminosity distance and red shift in a local
void~\cite{Tomita,Silk}. However, the determination of the sign of
the deceleration parameter $q$ depends on performing a spatial
volume averaging of physical scalar quantities, such as the time
evolution of the expansion parameter $\theta$ in the Raychoudhuri
equation~\cite{Moffat}.

Recently, Barausse, Matarrese and Riotto,~\cite{Matarrese,Pyne},
Kolb et al.~\cite{Kolb} and Notari~\cite{Notari} have perturbed a
homogeneous and isotropic FLRW universe to second order, and shown
that the luminosity distance-red shift relation implies a
non-vanishing cosmic variance of the deceleration parameter
implying an intrinsic theoretical uncertainty in its
determination. They uncover an infrared long-wavelength effect
that depends on the fluctuation spectrum $n_s$. For $n_s-1 \ll 0$,
a unit variance is obtained in second order perturbation that
generates a significant cosmic variance in the deceleration
parameter $q_0$ for cosmological perturbations with the largest
wavelengths. These results have been criticized by various
authors~\cite{Chung,Flanagan,Seljak,Wiltshire,Siegel,Giovannini,Rasanen}.

Several of these investigations rely on using specific gauge
conditions to carry out the perturbation calculations. Unless all
the calculations of physical quantities are performed within a
fully gauge invariant formalism, the choice of specific gauges
such as the synchronous gauge or the Poisson (Newtonian) gauge can
produce gauge dependent results. Moreover, some of the
calculations assume an FLRW background spacetime.
R\"as\"anen~\cite{Rasanen2} has considered the backreaction on a
dust dominated Lema\^{i}tre-Tolman-Bondi (LTB) background
spacetime~\cite{Lemaitre,Tolman,Bondi,Bonnor,Moffat2,Moffat3,Krasinski}.
Perturbation calculations fail in the late-time nonlinear regime
of large scale galaxy and void structure for red shifts $z\leq
25$.

We shall not assume that the universe is a global FLRW model.
Instead, we use an exact solution for the inhomogeneous
enhancement, based on field equations for an inhomogeneous
spherically symmetric distribution of matter, embedded in an
asymptotic FLRW universe, specializing to the matter dominated LTB
model. The assumption of spherical symmetry results in only one
inhomogeneous spatial dimension.

There have been recent reports of statistically significant
anomalies in the cosmic microwave background (CMB) compared to the
standard big bang
model~\cite{Hansen,Eriksen,Hajian,Gorski,Park,Ralston,Coles,Cruz,Vielva,Komatsu,Larson,
Mukherjee,Magueijo,Magueijo2,Oliveira,Tegmark,Copi,Starkman,Weeks,Dennis,Donoghue,Pisano}.
The WMAP data appear to reveal anomalies at the largest angular
scales $> 60^0$. The angular two-point correlation function is
suppressed at scales larger than $60^0$. When related to Fourier
space, the vanishing of the two-point correlation function at
large scales reveals an alignment of the quadrupole and octopole
moments. This could be due to cosmic variance. Oliveira-Costa et
al.,~\cite{Oliveira} studied the quadrupole and octopole data and
showed that both are planar and aligned with all maxima and minima
falling on a great circle on the sky. Similar results have been
reported by Schwarz et al.~\cite{Starkman} and Weeks~\cite{Weeks}.
It is not expected that these anomalies in the WMAP data are due
to instrumental failure. The CMB temperature fluctuations have
been found to be unevenly distributed between the southern and
northern hemispheres with a statistical significance $\sim
2-3\sigma$. Such effects could be produced by foreground
contamination. If, however, they are true cosmic effects, then
they would not agree with the standard predictions of inflation
theory and a significant fine-tuning of inflationary potentials
and parameters would be required to explain the
phenomenon~\cite{Contaldi}.

Schwarz et al., have applied a method assigning directions to the
$\ell$-th multipole using multipole vectors~\cite{Starkman}. They
found that the method reveals a high statistical significance
($99.9\%$) CL that the observed quadrupole and octupole are not in
agreement with a Gaussian random, statistically isotropic sky as
predicted generically by inflation. Surprisingly, they discovered
a strong correlation with the orientation of the ecliptic plane
(solar system and its motion) but no significant correlation with
the Milky Way. The motion of the solar system is related to the
measured CMB dipole. This effect could be due to a significant
systematic error in the WMAP data or that the largest scales of
the CMB sky are dominated by a local foreground. It could also be
due to cosmic variance and suggest that we are in a special place
in the universe at the present time. The cosmic variance for
Gaussian random variables is given by
\begin{equation}
\sigma(C_\ell)=\sqrt{\frac{2}{2\ell+1}}.
\end{equation}
It is a serious limitation for low multipoles that cannot be
avoided. If, on the other hand, it is due to a true cosmic effect,
or an absorbing or emitting source of microwave radiation in the
solar system, then this will have significant consequences for
interpreting cosmological data.

Normally, the known Doppler shifts due to the motion of Earth
round the Sun ($30\, {\rm km}\, {s}^{-1}$), the motion of the
solar system in the Galaxy ($\sim 250\,{\rm km}\,{s}^{-1}$) and
the motion of the Galaxy in the Local group of about $600\,{\rm
km}\,{s}^{-1}$ in the direction $l\sim 260\,^0, b\sim 30^0$ are
removed from the data. Any residual velocities observed would not
be expected to produce an alignment of the axes of lower
multipoles. An alternative possibility is the existence of
inhomogeneities on a scale approaching the scale of the visible
universe itself. The CMB data show that such inhomogeneities must
have a density contrast less than $10^{-4}$ over scales of order
$c/H_0$. The deviation of the observer from Hubble flow induced by
the perturbations give a variation in temperature of
amplitude~\cite{Sachs,Rees}:
\begin{equation}
\vline\frac{\delta T}{T}\vline\sim \vline
q_0\biggl(\frac{\delta\rho}{\rho}\biggr)\frac{a_s}{c/H_0}\vline,
\end{equation}
where $a_s$ is the scale of the perturbation and $q_0$ is the
deceleration parameter for $t=t_0$ and $z=0$. The gravitational
potential effects produce a fluctuation of order
\begin{equation}
\vline\frac{\delta T}{T}\vline\sim
\biggl(\frac{\delta\Phi}{\Phi}\biggr)
\biggl(\frac{\delta\rho}{\rho}\biggr)
\biggl(\frac{a_s}{c/H_0}\biggr)^2,
\end{equation}
where $\Phi$ is the gravitational potential. The angular scale of
the perturbation is of order $a_sH_0/c$ and at the surface of last
scattering it is
\begin{equation}
(\delta\rho/\rho)_r=(1+z_r)[a_{sr}(1+z_r)H_r(1+z_r)^{-3/2}]^2,
\end{equation}
where the subscript $r$ denotes the quantity at the surface of
last scattering (recombination).

We have approximately
\begin{equation}
\frac{G\delta M}{R}\sim v\delta v,
\end{equation}
for a perturbation of excess mass $\delta M$ with an induced
peculiar velocity $\delta v$ in the Hubble flow $v=H_0R$, where
$R$ is the distance from the center of the density perturbation.
This corresponds to a density contrast
\begin{equation}
\frac{\delta\rho}{\rho}\sim \vline\frac{1}{q_0}\frac{\delta
v}{H_0}\frac{R^2}{a_s^3}\vline.
\end{equation}

We shall investigate the interpretation of the red shift and
luminosity distance relations in the LTB model. We will also
determine the possible lack of complete smoothness of the
temperature fluctuations and correlation functions across the sky
predicted by the LTB model. The solution can be shown to have a
high degree of isotropy and homogeneity at large scales, but there
is a generic component of inhomogeneity and anisotropy that is
direction dependent and correlated with the location of the origin
in spherical polar coordinates.

\section{The Lema\^{i}tre-Tolman-Bondi Model}

The CMB temperature effects were first discussed by Sachs and
Wolfe~\cite{Sachs} and Kantowski~\cite{Kantowski} considering
small density contrasts over large scales for an Einstein-de
Sitter universe and that the observer is located in a
perturbation, or the perturbations were so distant that the
radiation passing through them occurred at last scattering. Other
situations were discussed by Rees and Sciama~\cite{Rees} and also
by Dyer~\cite{Dyer}. Raine and Thomas~\cite{Raine} analyzed the
situation in which the observer is situated near the edge of a
large-scale small amplitude density enhancement in an open
universe. We shall investigate in the following, a small amplitude
and large scale spherically symmetric density distribution with
the observer in a spatially flat universe.

The inhomogeneous line element in comoving coordinates can be
written as (see Appendix A):
\begin{equation}
ds^2=dt^2-R^{\prime 2}(t,r)f^{-2}(r)dr^2-R^2(t,r)d\Omega^2,
\end{equation}
where we choose units $G=c=1$, $d\Omega^2=d\theta^2+\sin^2\theta
d\phi^2$ and $f(r)$ is an arbitrary function of $r$ only. For the
matter dominated LTB model with zero pressure $p=0$ and zero
cosmological constant $\Lambda=0$, the Einstein field equations
demand that $R(t,r)$ satisfies
\begin{equation}
\label{Requation} 2R{\dot R}^2+2R(1-f^2)=F(r),
\end{equation}
with $F$ being an arbitrary function of class $C^2$, ${\dot
R}=\partial R/\partial t$, and $R^\prime=\partial R/\partial r$.
There exist three possible solutions depending on whether $f^2 <
1, = 1, > 1$ and they correspond to elliptic (closed), parabolic
(flat), and hyperbolic (open) cases, respectively.

The proper density of matter can be expressed as
\begin{equation}
\label{density} \rho=\frac{F'}{16\pi R'R^2}.
\end{equation}
By using (\ref{Requation}), we can solve (\ref{density}) to obtain
\begin{equation}
\Omega-1\equiv\frac{\rho}{\rho_c}-1=\frac{1}{3H^2_{\rm
eff}}\biggl(\frac{1-f^2}{R^2}-2\frac{f}{R}\frac{f'}{R'}\biggr),
\end{equation}
where
\begin{equation}
H^2_{\rm eff}=\frac{1}{3}(H^2_\perp+2H_\perp H_r).
\end{equation}
Here, we have defined two Hubble parameters $H_r(t,r)$ and
$H_\perp(t,r)$ for the local expansion of a spherically symmetric
density perturbation, corresponding to the radial and
perpendicular directions of expansion, respectively. We have
\begin{equation}
H_r=\frac{{\dot R}'}{R'},\quad H_\perp=\frac{{\dot R}}{R}.
\end{equation}

There exist three possibilities for the curvature of spacetime: 1)
$f^2
> 1$ open ($\Omega-1 < 0$), 2) $f^2=1$ flat ($\Omega-1=0$), $f^2 <
1$ closed ($\Omega-1 > 0$). We define a critical density in
analogy with the FLRW model for $f^2=1$:
\begin{equation}
8\pi\rho_c=\frac{{\dot R}^2}{R^2}+2\frac{{\dot R}}{R}\frac{{\dot
R}'}{R'}.
\end{equation}
This corresponds to the critical density for flat spatial sections
$t={\rm const}$. We can obtain the analogy of the Friedmann
equation for a spatially flat density perturbation:
\begin{equation}
H^2_{\rm eff}=\frac{8\pi\rho_c}{3}.
\end{equation}

The total mass of matter within comoving radius $r$ is
\begin{equation}
M(r)=\frac{1}{4}\int^r_0dr f^{-1}F'=4\pi\int^r_0dr\rho f^{-1}R'
R^2.
\end{equation}

Since the WMAP data~\cite{Spergel} shows that the universe is
spatially flat to within a few percent, we shall consider the
globally flat case $f^2=1$. The metric reduces to
\begin{equation}
\label{flatmetric} ds^2=dt^2-R^{'2}dr^2-R^2d\Omega^2,
\end{equation}
where
\begin{equation}
R(t,r)=r[t+\beta(r)]^{2/3},
\end{equation}
and $\beta(r)$ is an arbitrary function of $r$ of class
$C^2$~\cite{Bonnor}. The metric (\ref{flatmetric}) becomes
\begin{equation}
ds^2=dt^2-(t+\beta)^{4/3}(Y^2dr^2+r^2d\Omega^2),
\end{equation}
where
\begin{equation}
Y=1+\frac{2r\beta'}{3(t+\beta)},
\end{equation}
and
\begin{equation}
\rho=\frac{1}{6\pi(t+\beta)^2Y}.
\end{equation}

The arbitrary function $\beta(r)$ can be specified in terms of a
density on some spacelike hypersurface $t=t_0$. The metric and
density are singular on the two hypersurfaces $t+\beta=0$ and
$Y=0$, namely, $t_1=-\beta$ and $t_2=-\beta-2r\beta'/3$,
respectively. The model is only valid for $t > \Sigma(r)\equiv
{\rm Max}[t_1(r),t_2(r)]$, and the hypersurface $t(r)=\Sigma(r)$
defines the big-bang. However, our pressureless model requires
that the surface $t(r)=\Sigma(r)$ describes the surface on which
the universe becomes matter dominated (in the FLRW model this
occurs at $z\sim 10^4$). We observe that even in the spatially
flat LTB model, different parts of the universe can enter the
matter dominated era at different times.

For $\beta=0$ and in the limit $t\rightarrow\infty$ we obtain the
Einstein-de Sitter universe
\begin{equation}
ds^2=dt^2-a^2(t)(dr^2+r^2d\Omega^2),
\end{equation}
where $a(t)=t^{2/3}$. Thus, for $\beta=0$ we obtain the FLRW
model. Moreover, the expanding flat LTB model necessarily evolves
to the homogeneous and isotropic FLRW model for a non-vanishing
density, whatever the initial conditions.

\section{Paths of Light Rays in the Inhomogeneous Cosmology}

The luminosity distance between an observer at the origin of
coordinate system $t_0,0$ and the source at
$(t_e,r_e,\theta_e,\phi_e)$ is
\begin{equation}
d_L=\biggl(\frac{{\cal L}}{4\pi{\cal F}}\biggr)^{1/2}
=R(t_e,r_e)[1+z(t_e,r_e)]^2,
\end{equation}
where ${\cal L}$ is the absolute luminosity of the source, ${\cal
F}$ is the measured flux and $z(t_e,r_e)$ is the red shift (blue
shift) for a light ray emitted at $(t_e,r_e)$ and observed at
$(t_0,0)$.

For $\theta=\phi={\rm constant}$ and $f^2(r)=1$, we have for a
light ray travelling inwards towards the center and
$X(r,t)=R'(r,t)$:
\begin{equation}
\frac{dt}{dr}=-R'(r,t),
\end{equation}
where the sign is determined by the fact that the light ray
travels along the past light cone. Consider two rays emitted by a
source with a small time separation $\tau$. The equation of the
first ray is
\begin{equation}
t=T(r),
\end{equation}
while for the second ray, we have
\begin{equation}
t=T(r)+\tau(r).
\end{equation}
The equation of a ray and the rate of change of $\tau(r)$ along
the path is
\begin{equation}
\frac{dT(r)}{dr}=-R'[T(r),r],
\end{equation}
\begin{equation}
\frac{d\tau(r)}{dr}=-\tau(r){\dot R}'[T(r),r],
\end{equation}
where
\begin{equation}
{\dot R}'[T(r),r]={\dot R}'\vert_{r,T(r)}.
\end{equation}
Choosing $\tau(r_e)$ to be the period of a spectral line at $r_e$,
we get
\begin{equation}
\frac{\tau(0)}{\tau(r_e)}=\frac{\nu(r_e)}{\nu(0)}=1+z(r_e),\quad
z=0\quad {\rm for}\quad r_e=0.
\end{equation}

The red shift considered as a function of $r$ along the light cone
is determined by
\begin{equation}
\frac{dz}{dr}=(1+z){\dot R}'[T(r),r].
\end{equation}
For a light ray travelling from $(t_1,r_1)$ to $(t_0,0)$ the shift
$z_1$ is
\begin{equation}
\label{lightray} \ln(1+z_1)=-\int_0^{r_1} dr{\dot R}'[T(r),r].
\end{equation}
We can compare this to the result obtained from a FLRW universe
model by relabelling the radial coordinate $\bar r=R[T(r),r]$ and
choosing the solution $f^2(r)=1$, which corresponds for
$p=\Lambda=0$ to a flat, inhomogeneous matter dominated solution.
Moreover, we choose
\begin{equation}
a_1(r)=\dot R[T(r),r],\quad a_2(r)=\ddot R[T(r),r],...
\end{equation}
as functions of $r$ only. By expanding (\ref{lightray}), we obtain
\begin{equation}
\label{zequation}
\ln(1+z_1)=\int_0^{r_1}dr\frac{a_1'}{1-a_1}-\int_0^{r_1}dr\frac{a_1a_1'-a_2}
{1-a_1}=-\ln(1-a_1)-\int^{r_1}_0dr\frac{M'(r)}{(1-a_1)}.
\end{equation}
Two terms contribute to the cosmological red shift: the
contribution due to the expansion of the universe, and the shift
due to the difference between the potential energy per unit mass
at the source and at the observer. For the FLRW case, $M'(r)=0$
and there is no gravitational shift. The integral in
Eq.(\ref{zequation}) for small $r_1$ can be neglected, and
expanding the logarithms on both sides we find for small $r_e$ or
small $t_0-t_e$:
\begin{equation}
z(t_e,r_e)=H_\perp(t_e,r_e)d_L(t_e,r_e).
\end{equation}
The difference between this result and the one obtained from FLRW
is that it is {\it local} and depends on the angular Hubble
parameter $H_\perp=\dot R/R$ rather than on the FLRW Hubble
parameter $H_{FLRW}=\dot a/a$.

A formula for the luminosity distance in an exact FLRW universe is
given by~\cite{Turner}:
\begin{equation}
(d_L)_{FLRW}(z)=c(1+z)\int^z_0\frac{du}{H(u)}.
\end{equation}
This can be reexpressed as
\begin{equation}
(d_L)_{FLRW}(z)=c(1+z)\vert 1-\Omega_0\vert^{-1/2}H_0^{-1}S
\biggl[\vert
1-\Omega_0\vert^{1/2}H_0\int_0^z\frac{dz}{H(z)}\biggr],
\end{equation}
where $S(x)=\sin(x)$ for $\Omega_0 > 1$, $\sinh(x)$ for $\Omega_0
< 1$, and $x$ for $\Omega_0 =1$. For a flat universe
\begin{equation}
\label{decelerationq}
(d_L)_{FLRW}(z)=c(1+z)H_0^{-1}\int_0^zdu\exp\biggl[-\int_0^u[1+q(v(r))]
d\ln(1+v)\biggr],
\end{equation}
where $q$ denotes the deceleration parameter. This result only
depends on the assumption of a FLRW universe and does not depend
directly on solutions of Einstein's gravitational equations. In
our inhomogeneous cosmology, we must generalize the formula for
$d_L$:
\begin{equation}
\label{LTBdistance} (d_L)_{LTB}(z(r)) =c(1+z(r))H_{\perp
0}^{-1}\int_0^{z(r)}du\exp\biggl[-\int_0^u[1+q(v,r)]
d\ln(1+v)\biggr].
\end{equation}
Thus, we interpret the angular Hubble parameter
$H_\perp(t,r)={\dot R}(t,r)/R(t,r)$ as the Hubble parameter that
replaces the FLRW expression $H(t)={\dot a}(t)/a(t)$.

Let us expand $R(r,t)$ in a Taylor series
\begin{equation}
R(r,t)=R[r,t_0-(t_0-t)] =R(r,t_0)[1-(t_0-t)\frac{{\dot
R}(r,t_0)}{R(r,t_0)} +\frac{1}{2}(t_0-t)^2\frac{{\ddot
R}(r,t_0)}{R(r,t_0)} -...]
$$ $$
=R(r,t_0)[1-(t_0-t)H_{0\perp}-\frac{1}{2}(t_0-t)^2q_0(r,t_0)H_{0\perp}^2-...],
\end{equation}
where $t_0$ denotes the present epoch and $H_{0\perp}={\dot
R}(r,t_0)/R(r,t_0)$. Moreover, we have
\begin{equation}
q_0(r,t_0)=-\frac{{\ddot R}(r,t_0)R(r,t_0)}{{\dot R}^2(r,t_0)}.
\end{equation}
By substituting for ${\ddot R}$ from Eq.(\ref{inhomoF2}) (Appendix
A), we obtain
\begin{equation}
\label{inhomodeceleration}
q_0(r,t_0)=\frac{1}{3}+\frac{4\pi\rho_0(r,t_0)}{3H_{0\perp}^2(r,t_0)}
-\frac{\Lambda}{3H_{0\perp}^2(r,t_0)}-\frac{1}{3}\frac{H_{0r}(r,t_0)}{H_{0\perp}(r,t_0)},
\end{equation}
where $H_{0r}(r,t_0)={\dot R}'(r,t_0)/R'(r,t_0)$.

If we set $H_{0\perp}(r,t_0)=H_{0r}(r,t_0)=H_0$ where $H_0={\dot
a}(t_0)/a(t_0)$, then we obtain the spatially flat FLRW expression
for the deceleration parameter:
\begin{equation}
q_0=\frac{4\pi\rho_0}{3H_0^2}-\frac{\Lambda}{3H_0^2}=\frac{1}{2}\Omega_{0M}-\Omega_{0\Lambda}.
\end{equation}
We see from (\ref{inhomodeceleration}) that different observers
located in different causally disconnected parts of the sky will
observe different values for the deceleration parameter $q_0$,
depending upon their location and distance from the center of the
spherically symmetric inhomogeneous enhancement.

We must interpret the Hubble parameter and red shift measurements
in our model given a void on the scale of $\sim 100$ Mpc or a
large scale inhomogeneous enhancement of the scale of the Hubble
horizon $\sim c/H$. Observers at spatially separated locations in
the universe would interpret measurements of $z, H$ and $d_L$
differently from one another. We interpret the deceleration
parameter in the FLRW model as
\begin{equation}
q_0=-\frac{{\ddot a}(t_0)}{a(t_0)H_0^2},
\end{equation}
and from the SNe Ia supernovae observations arrive at a value $q_0
< 0$ corresponding to an accelerating universe. However, an
observer in the inhomogeneous model universe in a different, {\it
causally disconnected location in the universe} could observe a
different $q_0$ and luminosity distance $d_L$. Thus, averaging all
the local observations of $d_L$ and evaluations of $q_0$ can lead
to a cosmic variance on the determination of the deceleration
parameter. A significant generic, theoretical uncertainty (cosmic
variance) is generated in the determination of the deceleration
parameter $q_0$.

Luminosity distances are determined by supernovae measurements,
which cannot directly determine the instantaneous expansion rate
or deceleration rate. Thus, to determine the expansion history,
assumptions must be made about the evolution of $H(z)$ and $q(z)$.
We find that our inhomogeneous, spherically symmetric model of
large scale inhomogeneous enhancements, reaching out to the Hubble
horizon scale $c/H$, can lead to significant differences in the
local determinations of $H$ and $d_L$ and an observer independent
inference of the rate of expansion and deceleration of the
universe has a generic, built-in theoretical uncertainty. If the
universe were locally LTB for a matter dominated universe rather
than FLRW, then the Hubble parameter based on the observed LTB
values of $z$ determined through the relation (\ref{LTBdistance})
would be position and red shift dependent. The dependence of $H$
versus $z$ and $z$ versus $d_L$ were given
in~\cite{Moffat,Moffat2} for different evolving ages of the
universe. It was found that $H(z)$ strongly depends on the choice
of model for the FLRW case, and the choice of the deceleration
parameter $q_0$ in (\ref{decelerationq}) and (\ref{LTBdistance})
through which we interpret the observations. The calculations
in~\cite{Moffat,Moffat2} were based on the equation describing the
ratio of the local density to the density of the FLRW universe
taken along the past light cone for a void:
\begin{equation}
\Omega_{\rm void}[z(r)]\equiv
\frac{\rho[T(r),r]}{\rho_{FLRW}[T(r)]}=\frac{3F'(r)[T(r)+\beta_0]^2}
{8R'[T(r),r]R^2[T(r),r]},
\end{equation}
where $\beta_0$ is a constant.

\section{CMB Fluctuations and Inhomogeneity}

The main assumptions of the standard big-bang model are that the
universe is spatially flat and dominated in the matter era by cold
dark matter (CDM) and dark energy. The universe enters the matter
dominated era with scale invariant, Gaussian, adiabatic initial
fluctuations generated in the early universe by a period of
inflation. Through a hierarchical gravitational instability
process these uniform fluctuations started growing after
decoupling (surface of last scattering) to form the large scale
structure in the form of galaxies and clusters of galaxies that we
see today.

In spatially flat models, the density contrast is define by
\begin{equation}
\delta({\bf x})\equiv\frac{\delta\rho({\bf
x})}{{\bar\rho}}=\frac{\rho({\bf x})-\bar\rho}{\bar\rho},
\end{equation}
where $\bar\rho$ denotes the mean density. Its Fourier transform
is
\begin{equation}
\delta_k=\frac{1}{V}\int d^3x\delta({\bf x})\exp(i{\bf k}\cdot{\bf
x}).
\end{equation}
The power spectrum $\vert\delta_k\vert^2$ corresponds to
statistical random gaussian fluctuations with the rms density
fluctuations
\begin{equation}
\frac{\delta\rho}{\rho}=\langle\delta({\bf x})\delta({\bf
x})\rangle^{1/2}.
\end{equation}
For an isotropic power spectrum (i.e. depending on $k=\vert{\bf
k}\vert$ instead of ${\bf k}$):
\begin{equation}
\frac{\delta\rho}{\rho}=\frac{1}{V}\int^\infty_0
\frac{k^3\vert\delta_k\vert^2}{2\pi^2}\frac{dk}{k}.
\end{equation}
The autocorrelation function is defined by
\begin{equation}
\xi({\bf r})=\langle\delta({\bf x}+{\bf r})\delta({\bf x})\rangle,
\end{equation}
which is the Fourier transform of the power spectrum
$\vert\delta_k\vert^2$ and $\xi(0)=(\delta\rho/\rho)^2$.

Since our LTB model is confined to the matter dominated era, it is
reasonable to assume that to a first approximation in our
inhomogeneous flat model the density perturbations grow like
$\delta_{+}\equiv (\delta\rho/\rho)\propto t^{2/3}$ independently
of the scales. Moreover, we assume that the fluctuations entering
the matter dominated era on the surface $t(r)=\Sigma(r)$ have the
inflationary scale invariant behavior of the CDM FLRW model. In
the FLRW model, we have for the growing mode~\cite{Moffat2}:
\begin{equation}
\delta_{FLRW}(t)=\delta_{FLRW}(t_{eq})\biggl(\frac{t_{FLRW}}{t_{eq}}\biggr)^{2/3},
\end{equation}
where $t_{eq}$ and $t_{FLRW}$ denote the time of decoupling and
the time from the initial FLRW singularity to a given coordinate
value of time $t$, respectively. In the FLRW model these times are
the same everywhere in space.

In the LTB model for a spatially flat universe, we have
\begin{equation}
\delta_{LTB}(t,r)=\delta_{FLRW}(t_\Sigma(r),r)\biggl[\frac{t_{LTB}(r)}
{t_\Sigma(r)}\biggr]^{2/3},
\end{equation}
where now $t_{LTB}(r)=t-t_{\Sigma(r)}$ is the time from the
initial singularity $t>t_{\Sigma(r)}$ to a given coordinate time
$t$. For some time coordinate value $t$:
\begin{equation}
\label{fluctuations}
\delta_{LTB}(t,r)=\delta_{FLRW}(t)\biggl[\frac{t_{LTB}(r)}{t_{FLRW}}\biggr]^{2/3}.
\end{equation}
We have made the simplifying assumption:
\begin{equation}
\delta_{FLRW}(t_{\rm eq})=\delta_{LTB}(t_\Sigma(r),r)\quad {\rm
and}\quad t_{\rm eq}=t_\Sigma(r)\quad {\rm for\, all\, r}.
\end{equation}

For some spacelike hypersurface $t={\rm const}$, the gravitational
amplification of the primeval fluctuation perturbations depends on
the spatial position. The larger $t_{LTB}(r)$ for a given $r$, the
more developed structure we expect to see. Also, there can be a
different observable distribution of fluctuations in different
parts of the sky depending on $t_{LTB}(r)$ and the location of the
observer.

We can now write the two-point correlation function as
\begin{equation}
\xi_{LTB}(t_0,r)=\langle\delta_{LTB}(t_0,r)\delta_{LTB}(t_0,0)\rangle.
\end{equation}
The correction to the FLRW correlation function $\xi_{FLRW}$ is
given by
\begin{equation}
\xi_{LTB}(r)=\xi_{FLRW}(r)\biggl[\frac{t_{LTB}(r)}{t_{FLRW}}\biggr]^{2/3}.
\end{equation}

The CMB temperature anisotropies generated by scalar perturbations
of the FLRW spacetime metric can be written as
\begin{equation}
ds^2=a^2(\eta)[(1+2\Phi)d\eta^2-(1-2\Phi)d{\vec x}^2],
\end{equation}
where $a(\eta)$ is the FLRW scale factor and $\eta$ is the
conformal time. At the last scattering hypersurface, the
anisotropy contributes $\Phi/3$ sources. For a non-zero
$d\Phi/d\eta$ an additional integral contribution arises. for a
scale free primordial spectrum $\vert\Phi_k\vert^2=\vert
A^2/k^3\vert$ the CMB angular power spectrum amplitude can be
written~\cite{Contaldi}:
\begin{equation}
\langle\biggl(\frac{\delta T}{T}\biggr)^2\rangle_\ell
={36\pi^2}\biggl(\frac{2\ell+1}{\ell(\ell+1)}\biggr){A^2}K_\ell^2,
\end{equation}
where
\begin{equation}
K_\ell^2=2\ell(\ell+1)\int_0^\infty\frac{dk}{k}[j_\ell(k\eta_0)
+6\int_{\eta_r}^{\eta_0}d\eta
\frac{df_k}{d\eta}j_\ell(k(\eta_0-\eta))]^2,
\end{equation}
and $df_k/d\eta$ denotes $d\Phi_k/d\eta$ normalized to
$A/k^{3/2}$. A flat CDM model has $d\Phi/d\eta=0$ and
$K_\ell^2=1$, while a flat $\Lambda CDM$ has $K_\ell^2 > 1$ with
an amplification exceeding the Sachs-Wolfe plateau.

For the late time universe, it was argued by Contaldi et
al.~\cite{Contaldi} that the suppression of the large scale (low
$\ell$ moments) power spectrum could be explained by fine-tuning
the coefficients $K_\ell^2$ to be small with certain solutions of
late time $\Phi$. On the other hand, the amplitude of metric
perturbations at horizon crossing in inflationary models could
fine-tune $k^{3/2}\Phi_k$ to give a small contribution at large
angular scales. This would involve fine-tuning the shapes of
inflaton potentials in an ad hoc way.

In our LTB model, the growth of the angular power spectrum will
have the form
\begin{equation}
\label{powerspectrum} \langle\biggl(\frac{\delta
T}{T}\biggr)\rangle_\ell^{LTB}(t,r)=\langle\biggl(\frac{\delta
T}{T}\biggr)\rangle_\ell^{FLRW}(t)\biggl[\frac{t_{LTB}(r)}{t_\Sigma}\biggr]^{2/3}.
\end{equation}
We see that there will be a difference in the observed CMB
temperature fluctuations in different parts of the universe,
depending on the value of $r$ and the location of the observer.

\section{Isotropy of the Cosmic Microwave Background}

The standard description of cosmology is based on the FLRW model
and the cosmological principle -- the statement that the universe
is the same in all directions and locations. The assumption of
isotropy necessarily implies that the statistical properties of
the CMB should be the same in all directions on the sky. The FIRAS
experiment on the COBE satellite~\cite{Smoot,Bennett} demonstrated
that the mean temperature of the CMB is isotropic to a high
precision. However, the sensitivity of the WMAP
data~\cite{Spergel} can be used to test the isotropy of the
angular fluctuations in the CMB. Hansen, Banday and G\'orski
~\cite{Hansen} have studied the WMAP power spectrum extracted from
the CMB using patches at different directions on the sky. In the
lower angular range $\ell\sim 2-40$, they find preliminary
indications for non-zero differences in the power spectrum in the
northern and southern hemispheres oriented along the galactic
colatitude and longitude ($80^0, 57^0$), close to the ecliptic
pole. Eriksen et al.~\cite{Eriksen} and Park~\cite{Park} were the
first to discover a difference between the northern and southern
hemispheres in non-gaussianity tests. Similar results have been
found by Coles et al.~\cite{Coles}, Cruz et al.~\cite{Cruz},
Vielva et al.~\cite{Vielva}, Komatsu, Spergel and
Wandelt~\cite{Komatsu}, Larson and Wandelt~\cite{Larson}, Land and
Magueijo~\cite{Magueijo}, Oliveira et al.~\cite{Oliveira}. An
analysis of the variation in cosmological parameters associated
with the acoustical peak in the power spectrum was carried out by
Donoghue and Donoghue, to see whether there exists a correlation
between the height and location of the peak~\cite{Donoghue}. A
measurable dipole effect in the CMB would indicate that there
exists a spatial asymmetry. However, at present after the
subtraction of the dipole contribution from the WMAP data, there
is no indication of a statistically significant dipole
contribution.

The observed non-zero differences in the power spectrum in the
northern and southern hemispheres can be ascribed in our LTB model
as being due to a difference in $t_{LTB}(r)$ in the locations of
the northern and southern hemispheres, which will lead to a
correction factor described in Eqs.(\ref{fluctuations}) and
(\ref{powerspectrum}):
\begin{equation}
\Delta=\biggl(\frac{t_{LTB}(r_s)}{t_\Sigma}\biggr)-
\biggl(\frac{t_{LTB}(r_n)}{t_\Sigma}\biggr),
\end{equation}
where $r_s$ and $r_n$ denote the position locations of the
southern and northern hemispheres, respectively.

The claims about the detection of an asymmetric distribution of
large scale power in the CMB anisotropy as measured by the WMAP
satellite are found at large angular scales. For scales $\ell
> 40$ (corresponding to an angular scale $\sim 3-5^0$), the
distribution of the CMB fluctuations is consistent with the
hypothesis of isotropy and homogeneity, although there have been
indications of possible foreground contamination around the first
peak $\ell\sim 220$. However, for lower $\ell$ multipole values, a
strong difference between the northern and southern hemispheres
(for galactic and ecliptic frames of reference) was found. The CMB
power spectrum at large scales ($\ell=2-40$) is found to be
significantly much lower in the northern hemisphere than in the
southern hemisphere, which leads to a different estimate of the
cosmological parameters in these hemispheres.

When attention is focussed on three parameters to which the
analysis of the anisotropy and non-Gaussianity is most sensitive,
namely, the spectral index $n_s$, the amplitude of fluctuations
$A$ and the optical depth $\tau$, the preferred value for the
optical depth in the north is $\tau < 0.08$, whereas in the south
it is $\tau=0.24^{+0.06}_{-0.07}$ (68\% confidence level). The
latter result is inconsistent with $\tau=0$ at the $2\sigma$
level. The WMAP collaboration estimate $\tau=0.17$ could thus
originate in structure associated with the southern hemisphere. By
setting a prior on $\tau$, values of the spectral index $n_s$ are
found that are inconsistent between the opposite hemispheres.

Let us now assume that a mechanism is operative such that the
amplitude (\ref{powerspectrum}) is suppressed for large angular
scales in the sky for low multipoles. The observer is then able to
see small dipole, quadrupole and octopole contributions generated
by an inhomogeneous perturbation enhancement at scales $c/H$ and
the gravitational potential at the location of the enhancement. An
observer at $(r_0,t_0)$ measures a temperature $T=T_0$. At the
time of recombination $z\sim 10^3$, the temperature is $T_r\sim
10^3T_0$. The red shift varies somewhat with angle $\theta$
between the direction of a light ray emitted at $(r_0,t_0)$ and
absorbed at $(r_r,t_r)$ and a light ray directed toward the center
of the density perturbation. The apparent temperature of the CMB
as observed in the direction $\theta$ is given by
\begin{equation}
T_b(\theta)=\frac{T_r}{1+z(\theta)}=T_{\rm av}\frac{1+z_{\rm
av}}{1+z(\theta)},
\end{equation}
where $T_{\rm av}$ and $z_{\rm av}$ are the temperature and red
shift averaged over the whole sky and we have $T_{\rm av}\sim
T_0$.

The dipole, quadrupole and octopole moments, ${\cal D}, {\cal Q}$
and ${\cal O}$ are defined by~\cite{Piran}
\begin{equation}
{\cal D}\equiv \frac{1}{T_{\rm
av}}\int^\pi_0T_b(\theta)Y_{10}(\theta)\sin\theta d\theta,
\end{equation}
\begin{equation}
{\cal Q}\equiv \frac{1}{T_{\rm av}}\int_0^\pi
T_b(\theta)Y_{20}(\theta)\sin\theta d\theta,
\end{equation}
\begin{equation}
{\cal O}\equiv \frac{1}{T_{\rm av}}\int_0^\pi
T_b(\theta)Y_{30}(\theta)\sin\theta d\theta,
\end{equation}
where e.g. $Y_{10}=\sqrt{3/4\pi}\cos\theta$. An observer looking
toward the center of the spherically symmetric LTB density
perturbation will see an axially symmetric distribution. The
dipole ${\cal D}$, quadrupole ${\cal Q}$ and the octopole ${\cal
O}$ will be aligned in this plane, with the angle $\theta$
defining the trajectory of a light ray arriving to the observer
located at $(r_0,t_0)$. The dipole, quadrupole and octopole
contributions are produced by gradients of the gravitational
potential associated with the large scale perturbation
enhancement. The dipole contribution is small and undetected in
the WMAP data, while the quadrupole and octopole contributions
have been found to be planar and the planes are aligned to a
statistically anomalous degree~\cite{Oliveira,Starkman,Weeks}.

A possible explanation for the quadrupole and octopole anomaly is
that the cosmic topology has the form of a toroidal universe with
one small dimension of order one-half the horizontal scale, in the
direction towards Virgo (small universe model). Another
possibility is that the universe takes on the topological shape of
a dodecahedron~\cite{Luminet}. These possible explanations appear
to have been ruled out by Cornish and
collaborators~\cite{Cornish}.

\section{Conclusions}

We have used a cosmological model with the exact inhomogeneous LTB
solution for a matter dominated, large scale inhomogeneous
enhancement at Hubble horizon scale $c/H$ with $p=\Lambda=0$ to
demonstrate that observers located in different places in the
universe can differ significantly in their determinations of the
evolution of red shift and Hubble expansion rate. This will
significantly influence the theoretical interpretation of the
locally determined values of $H(z)$ and $d_L(z)$. Averaging over
all the observers' results will yield a generic cosmic variance in
the determination of the deceleration parameter. This in turn will
lead to a variance in the conclusion as to whether the universe is
undergoing an accelerating phase. An independent way to determine
luminosity distances to supernova and other distant objects in the
universe could settle the theoretical ambiguity.

The strong difference between the distribution of CMB fluctuations
in the northern and southern hemispheres discovered by analysis of
the WMAP power spectrum data for large scales with multipole
$\ell=2-40$ values, in the ecliptic and galactic frames of
reference, can be explained as a cosmological effect by a large
scale inhomogeneous enhancement, described by our exact
inhomogeneous solution of Einstein's field equations. The time
evolution of the universe as measured from the surface of last
scattering in the matter dominated LTB model can produce a
correction factor $(t_{LTB}(r)/t_\Sigma)^{2/3}$ for the
fluctuations that can differ significantly between the northern
and southern hemispheres. However, the uneven distribution of
fluctuations could be due to systematic effects in the WMAP data
or be due to foreground contamination. Hopefully, the WMAP2
analysis of the CMB data will decide whether or not the
non-Gaussian non-isotropic effect has a cosmological origin or is
due to some other non-cosmological mechanism.

The possible statistically significant alignment of the quadrupole
and octopole moments in the WMAP data can be explained, in our
large scale inhomogeneous model, by an off-center observer seeing
an axisymmetric alignment of the quadrupole and octopole moments
as the observer receives light rays from the center of the large
scale inhomogeneous enhancement.

\section{Appendix A}

For the sake of notational clarity, we write the FLRW line element
\begin{equation}
ds^2=dt^2-a^2(t)\biggl(\frac{dr^2}{1-kr^2}+r^2d\Omega^2\biggr),
\end{equation}
where $d\Omega^2=d\theta^2+\sin\theta^2d\phi^2$ and $k$ has its
usual interpretation.

Let us now consider the more general, spherically symmetric
inhomogeneous line element~\cite{Lemaitre,Tolman,Bondi}:
\begin{equation}
ds^2=dt^2-X^2(r,t)dr^2-R^2(r,t)d\Omega^2.
\end{equation}
The energy-momentum tensor ${T^\mu}_\nu$ takes the form
\begin{equation}
{T^\mu}_\nu=(\rho+p)u^\mu u_\nu -p{\delta^\mu}_\nu,
\end{equation}
where $u^\mu=dx^\mu/ds$ and, in general, the density
$\rho=\rho(r,t)$ and the pressure $p=p(r,t)$ depend on both $r$
and $t$. We have for comoving coordinates $u^0=1, u^i=0\quad
(i=1,2,3).$ Moreover, the total mass is
\begin{equation}
\label{mass} M(r)=4\pi\int_0^r{T^0}_0\sqrt{-g}=4\pi\int_0^r
dr\rho(r)XR^2,
\end{equation}
where $g={\rm Det}(g_{\mu\nu})=-X^2R^4\sin^2\theta$. From this
follows that
\begin{equation}
M'\equiv\frac{dM}{dr}=4\pi\rho XR^2.
\end{equation}
We have assumed, in (\ref{mass}), that geodesics of particles do
not intersect so that the total mass $M=M(r)$ depends on $r$
only~\cite{Bondi}.

The Christoffel symbols are
\begin{equation}
{\Gamma^\sigma}_{\mu\nu}=\frac{1}{2}g^{\sigma\alpha}(\partial_\mu
g_{\nu\alpha}+\partial_\nu g_{\mu\alpha}-\partial_\alpha
g_{\mu\nu}).
\end{equation}
The non-vanishing Christoffel symbols are
\begin{equation}
{\Gamma^1}_{11}=\frac{X'}{X},\quad {\Gamma^0}_{11}=X{\dot X},\quad
{\Gamma^1}_{01}=\frac{\dot X}{X},\quad
{\Gamma^2}_{02}={\Gamma^3}_{03}=\frac{\dot{R}}{R},
$$ $$
{\Gamma^0}_{22}=R\dot R,\quad {\Gamma^2}_{12}={\Gamma^3}_{13}
=\frac{R'}{R},\quad {\Gamma^1}_{22}=-\frac{RR'}{X^2},\quad
{\Gamma^0}_{33}=R \dot{R}\sin^2\theta,
$$ $$
{\Gamma^1}_{33}=-\frac{RR'}{X^2}\sin^2\theta,\quad
{\Gamma^2}_{33}=-\sin\theta\cos\theta,\quad
{\Gamma^3}_{23}=\cot\theta,
\end{equation}
where $X'=\partial X/\partial r$ and $\dot X=\partial X/\partial
t$. The Einstein gravitational equations are
\begin{equation}
G_{\mu\nu}+\Lambda g_{\mu\nu}=-8\pi T_{\mu\nu},
\end{equation}
where $G_{\mu\nu}= R_{\mu\nu}-\frac{1}{2}g_{\mu\nu}R$,
$R=g^{\mu\nu}R_{\mu\nu}$ and $\Lambda$ is the cosmological
constant. We have
\begin{equation}
\label{00} {G_0}^0\equiv 2\frac{\dot X\dot R}{XR}+\frac{1+{\dot
R}^2}{R^2}
-\frac{1}{X^2}\biggl(2\frac{R''}{R}+\frac{R'^2}{R^2}-2\frac{X'R'}{XR}\biggr)
=8\pi {T_0}^0+\Lambda=8\pi\rho+\Lambda,
\end{equation}
\begin{equation}
\label{11} {G_1}^1\equiv 2\frac{\ddot R}{R}+\frac{1+{\dot
R}^2}{R^2}-\frac{{R'}^2}{X^2R^2} =8\pi{T_1}^1+\Lambda=-8\pi
p+\Lambda,
\end{equation}
\begin{equation}
\label{22-33} {G_2}^2\equiv\frac{\ddot X}{X}+\frac{\ddot
R}{R}+\frac{{\dot X}{\dot
R}}{XR}-\frac{1}{X^2}\biggl(\frac{R''}{R}-\frac{X'R'}{XR}\biggr)
=8\pi{T_2}^2+\Lambda=8\pi{T_3}^3+\Lambda=-8\pi p +\Lambda,
\end{equation}
\begin{equation}
\label{10} {G_1}^0\equiv -2\biggl(\frac{\dot R'}{R}-\frac{\dot X
R'}{XR}\biggr) =8\pi {T_1}^0=-8\pi X^2{T_0}^1=0.
\end{equation}

From Eq. (\ref{10}), we find that
\begin{equation}
X(r,t)=\frac{1}{f(r)}R'(r,t).
\end{equation}
For an isotropic pressure ${T_1}^1={T_2}^2={T_3}^3$ and the
pressure $p=p(t)$ only depends on the time $t$.

We now obtain the two equations
\begin{equation}
\frac{{\dot R}^2}{R^2}+2\frac{{\dot R}'}{R'}\frac{{\dot
R}}{R}+\frac{1}{R^2}(1-f^2)-2\frac{ff'}{R'R}=8\pi\rho+\Lambda,
\end{equation}
\begin{equation}
\frac{\ddot R}{R}+\frac{1}{3}\frac{\dot{R}^2}{R^2}
+\frac{1}{3}\frac{1}{R^2}(1-f^2)-\frac{1}{3}\frac{\dot
R'}{R'}\frac{{\dot R}}{R} +\frac{1}{3}\frac{ff'}{R'R}
=-\frac{4\pi}{3}(\rho+3p)+\frac{1}{3}\Lambda.
\end{equation}

We obtain for $f^2=1$ the result
\begin{equation}
\label{inhomoF} \frac{{\dot R}^2}{R^2}+2\frac{\dot R}{R}\frac{\dot
R'}{R'}=8\pi\rho+\Lambda,
\end{equation},
\begin{equation}
\label{inhomoF2} \frac{\ddot {R}}{R}+\frac{1}{3}\frac{{\dot
R}^2}{R^2}-\frac{1}{3}\frac{{\dot R}'}{R'}
\frac{\dot{R}}{R}=-\frac{4\pi}{3}(\rho+3p)+\frac{1}{3}\Lambda.
\end{equation}
By using the notation $H_\perp=\dot{R}/R$ and $H_r=\dot{R}'/R'$,
we can write (\ref{inhomoF}) and (\ref{inhomoF2}) as
\begin{equation}
H^2_\perp+2H_\perp H_r=8\pi\rho+\Lambda,
\end{equation}
\begin{equation}
\frac{\ddot{R}}{R}+\frac{1}{3}H_\perp^2-\frac{1}{3}H_r
H_\perp=-\frac{4\pi}{3}(\rho+3p)+\frac{1}{3}\Lambda.
\end{equation}
For $H_\perp(r,t)=H_r(r,t)=H(t)=\dot{a}/a$ and $R(r,t)=a(t)$, we
obtain the Friedmann equations of FLRW for a spatially flat
universe:
\begin{equation}
H^2=\frac{8\pi\rho}{3}+\frac{1}{3}\Lambda.
\end{equation}
\begin{equation}
\frac{\ddot{a}}{a}=-\frac{4\pi}{3}(\rho+3p)+\frac{1}{3}\Lambda,
\end{equation}
where $\rho=\rho(t)$ and $p=p(t)$.

\vskip 0.2 true in {\bf Acknowledgments} \vskip 0.2 true in This
work was supported by the Natural Sciences and Engineering
Research Council of Canada. I thank Martin Green, Joao Magueijo,
Constantinos Skordis and Glen Starkman for helpful
discussions.\vskip 0.5 true in

\end{document}